\documentclass{aa}
\usepackage{xspace}
\usepackage{natbib}
\usepackage{graphicx}
\usepackage{amssymb}
\usepackage{amsmath}
\usepackage{url}
\usepackage{txfonts}
\usepackage{rotating}
\usepackage{microtype}

\bibpunct{(}{)}{;}{a}{}{,} 

\newcommand{\integral}{\textsl{INTEGRAL}\xspace}

\newcommand{\ca}{\mbox{$\sim$}}
\newcommand{\err}[2]{\ensuremath{^{+#1}_{-#2}}\xspace}
\newcommand{\Rsun}{\ensuremath{R_\odot}\xspace}
\newcommand{\Msun}{\ensuremath{M_\odot}\xspace}

\newcommand{\xte}{\textsl{RXTE}\xspace}

\newcommand{\asm}{{ASM}\xspace}

\newcommand{\nh}{\ensuremath{N_{\text{H}}}\xspace}
\newcommand{\redchi}{\ensuremath{\chi^2_{{\text{red}}\xspace}}}

\newcommand{\vela}{\mbox{Vela~X-1}\xspace}
\newcommand{\ecut}{\ensuremath{E_{\text{cut}}}\xspace}
\newcommand{\efold}{\ensuremath{E_{\text{F}}}\xspace}

\newcommand{\hd}{HD\,77581\xspace}
\newcommand{\cps}{\ensuremath{\text{counts}\,\text{s}^{-1}}\xspace}

\begin{document}
\title{High variability in \vela: giant flares and off states}

\author{\mbox{Ingo Kreykenbohm\inst{1,2,3,4}} \and 
  \mbox{J\"orn Wilms\inst{1,2}} \and
  \mbox{Peter Kretschmar\inst{5}} \and
  \mbox{Jos\'e Miguel Torrej{\'o}n\inst{6,7}} \and
  \mbox{Katja Pottschmidt\inst{8,9}} \and 
  \mbox{Manfred Hanke\inst{1,2}} \and
  \mbox{Andrea Santangelo\inst{3}} \and
  \mbox{Carlo Ferrigno\inst{3,4,10}} \and
  \mbox{R\"udiger Staubert\inst{3}} }
\offprints{I. Kreykenbohm,\\ e-mail: {Ingo.Kreykenbohm@sternwarte.uni-erlangen.de}}

\institute{Dr.  Karl Remeis-Sternwarte Bamberg, 
  Sternwartstrasse~7, 96049~Bamberg, Germany \and Erlangen Centre for
  Astroparticle Physics (ECAP), Erwin-Rommel-Str. 1, 91058~Erlangen,
  Germany \and Kepler Center for Astro and Particle Physics, Institut
  f\"ur Astronomie und Astrophysik, Sand 1, 72076 T\"ubingen, Germany
  \and \textsl{INTEGRAL} Science Data Centre, 16 ch.\ d'\'Ecogia, 1290
  Versoix, Switzerland \and European Space Agency, European Space
  Astronomy Centre, Villafranca del Castillo, P.O. Box 78, 28691
  Villanueva de la Ca{\~ n}ada, Madrid, Spain \and Departamento de
  F\'isica, Ingenier\'ia de Sistemas y Teor\'ia de la Se\~nal, Escuela
  Polit\'enica Superior, Universidad de Alicante, Ap.\ 99, 03080
  Alicante, Spain \and Kavli Institute for Astrophysics and Space
  Research, Massachusetts Institute of Technology, Cambridge, MA 02139, USA
  \and CRESST, University of Maryland, Baltimore County, 1000 Hilltop
  Circle, Baltimore, MD\ 21250, USA \and NASA Goddard Space Flight
  Center, Astrophysics Science Division, Code 661, Greenbelt, MD\ 20771,
  USA \and IASF--INAF, via Ugo la Malfa 153, 90136 Palermo, Italy
}

\date {Received: 14 April 2008 / Accepted: 2 August 2008}

\abstract{} {We investigate the spectral and temporal behavior of the
  high mass X-ray binary \vela during a phase of high activity, with
  special focus on the observed giant flares and off states.}
{\integral observed \vela in a long almost uninterrupted observation
  for two weeks in 2003 Nov/Dec. The data were analyzed with OSA~7.0
  and FTOOLS 6.2. We derive the pulse period, light curves, spectra,
  hardness ratios, and hardness intensity diagrams, and study the
  eclipse.}  {In addition to an already high activity level, \vela
  exhibited several intense flares, the brightest ones reaching a
  maximum intensity of more than 5\,Crab in the 20--40\,keV band and
  several off states where the source was no longer detected by
  \integral. We determine the pulse period to be
  $283.5320\pm0.0002$\,s, which is stable throughout the entire
  observation.  Analyzing the eclipses provided an improvement in the
  ephemeris. Spectral analysis of the flares indicates that there
  appear to be two types of flares: relatively brief flares, which can
  be extremely intense and show spectral softening, in contrast to
  high intensity states, which are longer and show no softening. }
{Both flares and off states are interpreted as being due to a strongly
  structured wind of the optical companion. When \vela encounters a
  cavity with strongly reduced density, the flux will drop triggering
  the onset of the propeller effect, which inhibits further accretion,
  giving rise to off states. The sudden decrease in the density of the
  material required to trigger the propeller effect in \vela is of the
  same order as predicted by theoretical papers about the densities in
  OB star winds. A similarly structured wind can produce giant
  flares when \vela encounters a dense blob in the wind. }


\maketitle

\section{Introduction}
\label{sect:intro}
\object{Vela X-1} (4U\,0900$-$40) is an eclipsing high mass X-ray
binary (HMXB) consisting of the B0.5Ib super giant \object{HD\,77581}
and a neutron star with an orbital period of 8.964\,days
\citep{kerkwijk95a} at a distance of $\sim$2.0\,kpc
\citep{nagase89a}. The optical companion has a mass of $\sim$23\,\Msun
and a radius of $\sim$30\,\Rsun \citep{kerkwijk95a}.  Due to the small
separation of the binary system with an orbital radius of just
$1.7\,{\rm R_\star}$, the massive 1.9\,\Msun neutron star \citep[\vela is
the most massive compact object known to be a neutron
star;][]{quaintrell03a,barziv01a} is deeply embedded in the dense
stellar wind of its optical companion \hd \citep[$\dot M_\star = 4
\times 10^{-6}$\,\Msun$\text{yr}^{-1}$;][]{nagase86a}. X-ray lines
indicate that this wind is inhomogeneous with many dense clumps
\citep{oskinova08a} embedded in a far thinner, highly
ionized component \citep{sako99a}.

The neutron star revolves with a long spin period of $\sim$283\,s
\citep{rappaport75a,mcclintock76a}.  Both the spin period and spin
period derivative have changed erratically since their first measurements,
as expected for a wind-accreting system. The evolution of the spin
period is most appropriately described by a random walk model
\citep{tsunemi89a,ziolkowski85a}. Although the source exhibits strong
pulse-to-pulse variations, a pulse-profile folded over several pulse
periods shows remarkable stability \citep[for 10 pulses or
more;][]{staubert80a}, even over decades \citep{raubenheimer90a}. At
energies below 5\,keV, the pulse-profile consists of a complex
five-peaked structure, which transforms at energies above 20\,keV
into a simple double-peaked pulse-profile \citep{staubert80a} where
the two peaks are thought to be due to the two accreting magnetic
poles of the neutron star.

With \ca$4 \times 10^{36}\,\text{erg\, s}^{-1}$, the X-ray luminosity
of \vela is typical of a high mass X-ray binary.  Observations in the
past, however, have shown that the source is variable with observed
flux reductions to less than 10\% of its normal value \citep[off
states;][]{kreykenbohm99a,kretschmar99a,inoue84a}, while periods of
increased activity have also been observed during which the flux
increases within an hour to a multiple of the previous value, reaching
peak flux levels close to 1\,Crab
\citep{kreykenbohm99a,haberl90a,kendziorra89a}. In this respect, \vela
is similar to sources such as \object{4U\,1700$-$377} and
\object{4U\,1907$+$09}, for which low luminosity states and flares
have also been observed, as is rather typical for wind-accreting
systems \citep[see e.g.][]{fritz06a,vandermeer05a,zand97a,haberl89a}.
Although \vela is a well studied object, only observations by
\integral revealed that the flares in \vela can be brighter than
previously anticipated \citep{staubert04a,krivonos03a}. While the
flaring activity is thought to be due to a strongly increased
accretion rate, $\dot{M}$, the origin of the $\dot{M}$ variations is
unknown.

The phase averaged X-ray spectrum of \vela was usually modeled
with a power law modified at higher energies by an exponential cutoff
\citep{tanaka86a,white83a} or with the Negative Positive EXponential
\citep[NPEX-model;][]{mihara95a}.  The spectrum was found to be modified by
strongly orbital-phase-dependent photoelectric absorption at lower
energies due to the dense stellar wind and an accretion wake trailing
the neutron star
\citep{goldstein04a,kreykenbohm99a,feldmeier96a,haberl90a}.  At
6.4\,keV, an iron fluorescence line and occasionally an iron edge at
7.27\,keV \citep{nagase86a} were observed in the X-ray spectrum. At
higher energies, cyclotron resonant scattering features (CRSFs)
between 25 and 32\,keV \citep{makishima92a,choi96a,kreykenbohm02b}
and at \ca55\,keV
\citep{kendziorra92a,orlandini97c,kreykenbohm99a,labarbera03a,attie04a}
were present, although the interpretation of the 25\,keV feature is
still sometimes debated \citep{orlandini05a}.

The remainder of this paper is structured as follows. In
Sect.~\ref{sect:data}, the data and software used are described.
Section~\ref{sect:analysis} describes first the temporal analysis of
the data, i.e. light curves, determination of the pulse period,
quasi-periodic modulations, and then the analysis of the eclipse and
the spectral analysis. The results are discussed in
Sect.~\ref{sect:discussion} and a summary is presented in
Sect.~\ref{sect:summary}.

\section{Data}
\label{sect:data}
\subsection{Instrument and Software}
The \integral observatory \citep{winkler03b} is in a highly eccentric
orbit with a period of $71^{\text{h}}49^{\text{m}}$, ideal for long
uninterrupted observations with a low X-ray background. Due to the
high eccentricity of the orbit the perigee passage (when \integral is
inside the radiation belts) has a duration of only \ca8\,h, which
minimizes the time when no science observations are possible due to
the high radiation background.

\integral has four science instruments, which provide coverage from
3\,keV up to 10\,MeV as well as in the optical: the imager IBIS/ISGRI
\citep[20\,keV to 800\,keV]{ubertini03a} with moderate energy
resolution and a large effective area, the spectrometer SPI
\citep[20\,keV to 10\,MeV;][]{vedrenne03a} for the analysis of nuclear
lines, the X-ray monitor JEM-X \citep[3\,keV to 35\,keV]{lund03a}, and
the optical monitor OMC \citep{mas-hesse03a}. All high-energy
instruments are coded-mask telescopes \citep[see e.g.][for a review of
this technique]{zand92a}. To improve the imaging quality, the
satellite performs raster observations of the vicinity of an X-ray
source, retaining the target in the field-of-view of ISGRI. Due to
this ``dithering'' strategy the off-axis angle of the target source
changes significantly during the observation. When the source is more
than \ca4\fdg{5} away from the pointing direction, it is in the
partially coded field-of-view of ISGRI. With increasing distance from
the pointing direction, the coding factor decreases, causing increased
uncertainties in the images, flux values, and spectra.  Individual
pointings made during these dithering observations are called science
windows (SCWs). They have typical durations of 1800\,s, 2200\,s, or
3600\,s.  These SCWs are then associated with \integral revolutions,
i.e. complete orbits of the \integral satellite around the Earth.

\begin{figure}
\centerline{\includegraphics[width=\columnwidth]{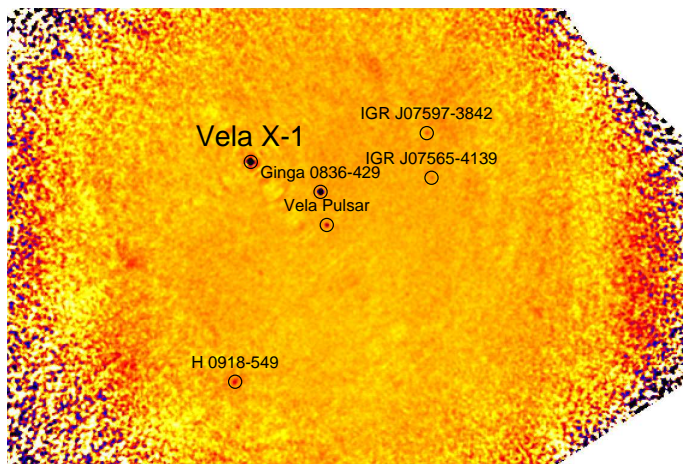}}
\caption{Intensity mosaic image using ISGRI data of the Vela region.
  All five revolutions from Rev.~137 to Rev.~141 have been used for
  this mosaic.  \vela is by far the most significant source. Several
  more sources are also detected.  Note that the times of the eclipses
  of \vela (see Fig.~\ref{fig:lc}) have been excluded from this
  mosaic, resulting in an exposure of \ca960\,ksec.  The noisy rim of
  the mosaic is due to the low coding factor in the outermost
  part of the partially coded field-of-view. }
\label{fig:image}
\end{figure}

To prepare the \integral data for analysis, we used the Offline
Science Analysis Software (OSA), version 7.0, and its associated
calibration files. In particular, we make extensive use of the tool
\textsl{ii\_light}. We carefully checked the behavior of
\textsl{ii\_light} (see Appendix~\ref{app_ii}), because the IBIS
cookbook\footnote{available at\\
  \url{http://isdc.unige.ch/?Support+documents}.}  cautions that
\textsl{ii\_light} should only be used to analyze the timing behavior
within a given science window.  For further analysis,
we used HEADAS release 6.2. Spectral fitting was done with
\textsl{XSPEC} 11.3.2ad \citep{dorman01a,arnaud96a}.

\subsection{Data}
\label{data}
\begin{figure*}[ht]
\centerline{\includegraphics[width=0.965\textwidth]{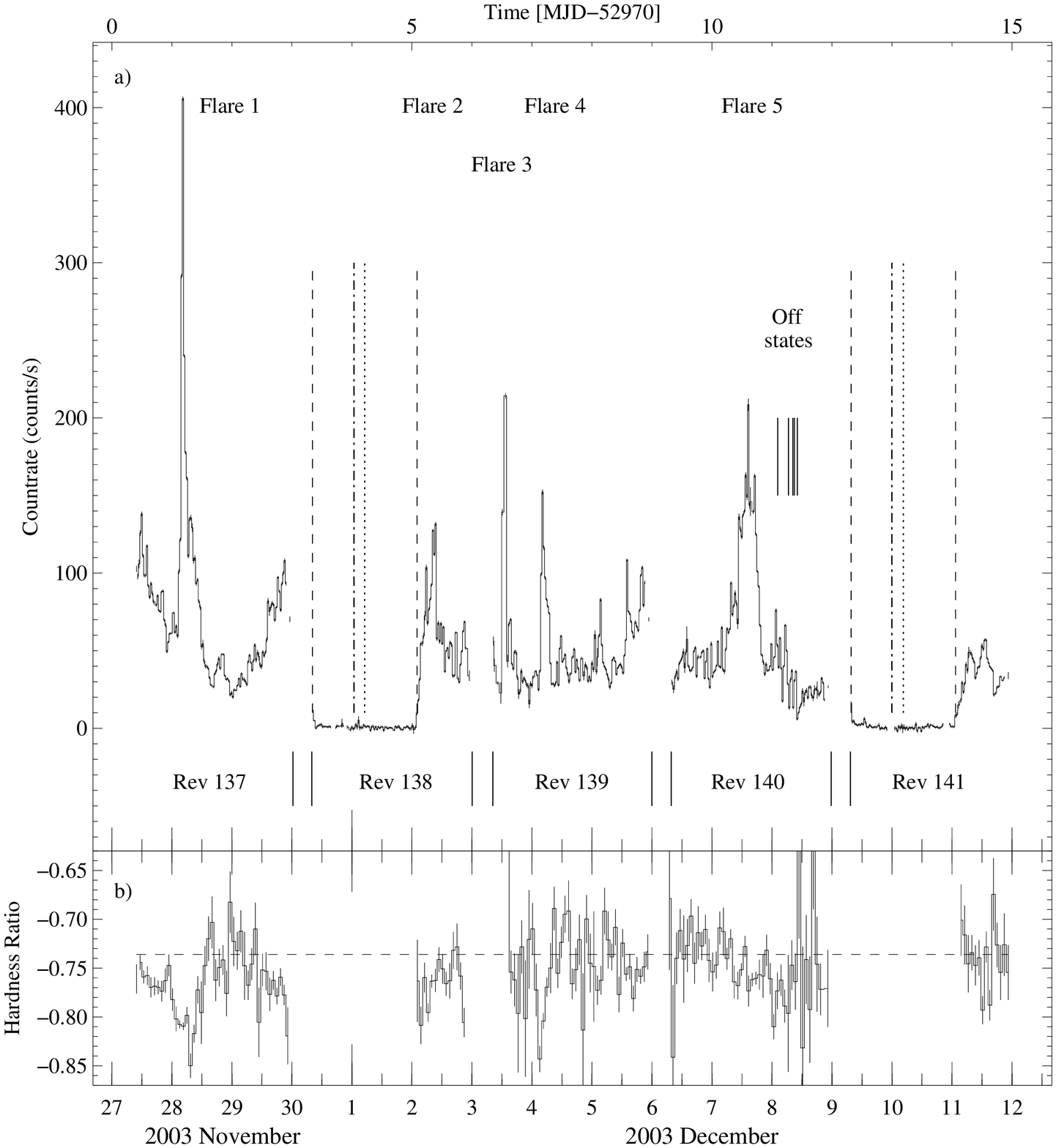}}
\vfill
\caption{Variability of \vela for the complete Vela region observation
  from Revolution 137 to 141. \textbf{a} ISGRI 20--40\,keV light curve
  (time resolution 1\,SCW, i.e. $\sim$1800\,s) and \textbf{b}
  20--30\,keV vs.\ 40--60\,keV hardness ratio as defined by
  Eq.~\ref{eq:hardness} (rebinned by a factor of $\sim$3 with respect
  to the light curve). Labels indicate the revolution number. 
  Short vertical lines below the X-ray light curve show
  \integral's perigee passages, during which the instruments are
  switched off.  The long dashed vertical lines show the
  ingress and egress times. The dotted vertical line
  indicates the derived eclipse center (see also
  Table~\ref{tab:ephemeris}), while the dash-dotted line indicates the
  time of mean longitude $T_{90}$ based on the ephemeris from
  \citet{nagase89a}. Note that the offset of the newly derived
  $T_{90}$ (see Table~\ref{tab:ephemeris}) in comparison to that of
  \citet{nagase89a} is too small to be visible in this figure.  See
  text for further discussion.  }
\label{fig:lc}
\end{figure*}

As part of the AO1 core program \citep{winkler01a}, \integral observed
the Vela region (see Fig.~\ref{fig:image}) continuously for five
consecutive \integral revolutions from the beginning of revolution~137
(JD\,2452970.86) until the end of revolution~141 (JD\,2452970.86)
resulting in approximately 1\,Msec of data (see
Fig.~\ref{fig:lc}). The observation was performed in a $5\times5$
dithering pattern with stable pointings $2^\circ$ apart.

We chose to use \emph{all} available science windows from Rev.~137 to
Rev.~141 to be able to derive a contiguous light curve with as few
interruptions as possible (data gaps due to the perigee passage of the
satellite are obviously unavoidable; see Fig.~\ref{fig:lc}). Since
\vela is a bright source, the OSA software has no problem in detecting
the source and determining its flux level accurately even when the
source is at an off-axis angle of more than $14^\circ$; in any case,
fewer than 5\% of the pointings had an off-axis angle larger than
$14^\circ$. For studying the timing behavior on timescales short
compared with a SCW, i.e. period determination and search for QPOs
(see Sect.~\ref{sect:period}), the absolute flux is not important and
the temporal properties are unaffected by a non-optimal off-axis flux
correction.

Figure~\ref{fig:image} shows the image of the Vela region from these
observations. While \vela is by far the brightest source in the
field-of-view of ISGRI, we also detect 4U\,0836$-$429 as a very
prominent source reaching about a third of the average intensity of
\vela, and the relatively weak sources \object{H\,0918$-$5459}, the
\object{Vela~Pulsar}, and two sources first reported by \integral
\citep{denhartog04a,sazonov05a}. Since the two brightest sources \vela
and \object{4U\,0836$-$429} are well separated (about 6\fdg{7}) and
all the weaker sources are even more distant, contamination of the
spectrum of \vela due to the presence of the other sources is of no
concern. Data from JEM-X and SPI have not been used in this analysis
due to the far smaller field-of-view of JEM-X and since \vela is
off-center in the observed field (\vela was only within the fully
coded field-of-view of JEM-X for less than ten out of the $\sim$550
individual pointings).  The SPI instrument, on the other hand,
provides a high spectral resolution, although, due to its low
effective area, it is not possible to study data on timescales of
seconds as required here.

\section{Data analysis}
\label{sect:analysis}
\subsection{Light curves and flux}
\label{sect:lc}
\vela was found in a strongly variable state during the Nov/Dec
2003 observation by \integral. While periods of increased activity
were observed before \citep{kreykenbohm99a,haberl94a}, the behavior
found in this observation \citep[see also][]{staubert04a} is indeed
extreme.

Most prominently, on 2003 November 28 (JD\,2452971.67), \integral
observed the brightest flare ever seen from \vela \citep[designated
flare~1; see Fig.~\ref{fig:lc} and also][]{krivonos03a}. During the
flare, the 20--40\,keV count rate increased from a SCW averaged
pre-flare value of \ca55\,\cps ($\sim$300\,mCrab, or $1.6\times
10^{-9}\,\text{erg}\,\text{s}^{-1}$) by a factor of more than seven to
405\,\cps (2.3\,Crab) within only 90\,minutes -- normal flaring
activity reaches peak flux values not higher than 1\,Crab \citep[see
e.g.][]{kreykenbohm99a}. \vela was therefore more than ten times
brighter than on average (see Table~\ref{tab:increase}) which implies
that flare~1 is a giant flare.
The average flux values for \vela in Table~\ref{tab:increase} were
derived from a spectrum with 400\,ksec of exposure obtained between
revolution 81 and 89 in June/July 2003, when \vela was in a normal
state. In the following, we designate a flare with a peak flux of more
than 2 Crab as a giant flare, as opposed to normal flares, which do
not reach this flux level. These flares reach their peak
rapidly, i.e.  $T_\text{rise}/T_\text{total} < 0.3$, where
$T_\text{rise}$ is the time from the onset of the flare to the
peak. While flare~1 was detected in all energy bands, it was most
pronounced in the 20--30\,keV band.  As stated above, \vela was never
in the field-of-view of JEM-X during the flare, such that coverage at
even lower energies is unavailable.

\begin{figure}
\includegraphics[width=\columnwidth]{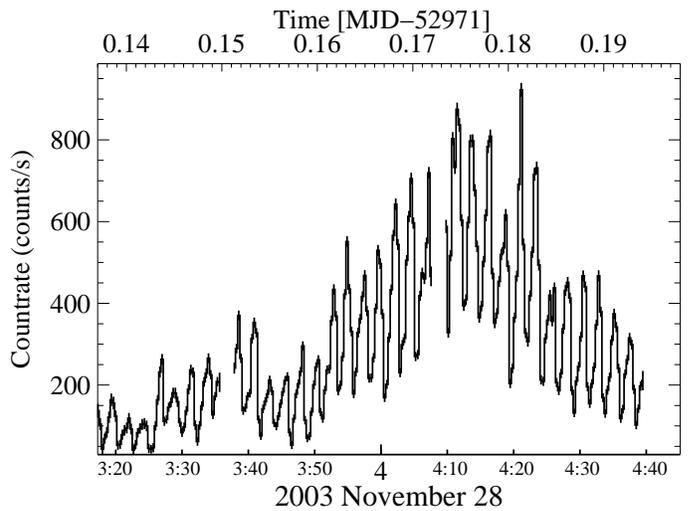}
\caption{Close up of the light curve with a time resolution of 20\,s
  of the peak of giant flare~1 on 2003 November 28. The peak is
  reached at MJD\,52971.18 with 923\,\cps (corresponding to 5.2\,Crab)
  in the 20 to 40\,keV band.  This value is significantly higher than
  the count rate averaged over one science window, since the count rate
  is not constant. Furthermore, the source is strongly pulsating,
  exhibiting the well known double pulse.}
\label{fig:flare_lc}
\end{figure}

After the peak (duration about half an hour, see
Fig.~\ref{fig:flare_lc}), the flare decayed quickly in less than 2\,h
to an intensity level of $<$1\,Crab and within \ca11\,h to a SCW
averaged count rate of \ca35\,\cps (200\,mCrab), somewhat lower than
before the onset of the flare (see Fig.~\ref{fig:lc}). The total
energy released between 20\,keV and 40\,keV amounts to
$1.15\times10^{41}$\,ergs.  We emphasize that these fluxes and those
given in Fig.~\ref{fig:lc} and Table~\ref{tab:increase} are SCW
averaged fluxes. These values average over fluctuations on shorter
timescales such as pulsations, during which the source sometimes
reached significantly higher intensities (see below and
Fig.~\ref{fig:flare_lc}).

\begin{table}
  \caption{Flux values of \vela in science window 013700420010, during which giant
    flare~1 reached its maximum. The increase indicates the factor by which \vela was
    brighter during the flare than during its normal state. The flux
    values are averages for the entire science window.
 For the peak
    flux reached in the pulses, see Table~\ref{tab:flares}.}
\label{tab:increase}
\begin{tabular}{r@{ \ }c@{ }r@{ \ }rrcc}
\hline
\hline
\multicolumn{3}{c}{energy}  & Flux (normal) & Flux (flare) &
Flux (flare) & increase \\
\multicolumn{3}{c}{[keV]}   & \multicolumn{2}{c}{[ $10^{-10}$ ergs cm$^{-2}$ s$^{-1}$ ]} & [Crab] & \\
\hline
20 &--&  30 & 10.7 & 136.3 & \phantom{$<$}2.8 & 13 \\
30 &--&  40 &  5.6 &  58.6 & \phantom{$<$}1.9 & 10 \\
40 &--&  50 &  1.9 &  16.6 & \phantom{$<$}0.7 & \phantom{1}9 \\
50 &--&  60 &  0.6 &   4.5 & \phantom{$<$}0.3 & \phantom{1}8 \\
60 &--&  80 &  0.7 &   2.6 & \phantom{$<$}0.1 & \phantom{1}4 \\
80 &--& 100 &  0.1 &   0.6 & $<$0.1& \phantom{1}6 \\
\hline
\end{tabular}
\end{table}

\begin{table}
  \caption{Overview of the observed flares. See Fig~\ref{fig:lc} for the
    numbering of the flares. The time is the onset of the flare. To
    obtain the peak fluxes $F_\text{peak}$, a light curve with a time resolution of
    20\,s was used. $T_\text{rise}$ is the time from the onset of
      the flare to the peak, while $T_\text{total}$ is the duration of the flare.}
\label{tab:flares}
\begin{tabular}{c@{ }c@{ }r@{ }c@{ }cp{0.3\columnwidth}}
  \hline
  \hline
  Flare & Time & Duration & $F_\text{peak}$ &
  $\frac{T_\text{rise}}{T_\text{total}}$ & Remarks\rule{0pt}{1.1em} \\
  & [MJD] & \multicolumn{1}{c}{[s]} & \multicolumn{1}{c}{[Crab]} & & \\
  \hline
  1 & 52971.15 & 11\,200 & 5.2 & 0.15 & giant flare, spectral softening\\
  2 & 52975.34 &  5\,200 & 2.6 & 0.83 & no spectral change\\
  3 & 52976.50 &  1\,800 & 5.3 & 0.28 & giant flare, very short\\
  4 & 52977.15 & 12\,900 & 1.9 & 0.13 & spectral softening \\
  5 & 52980.31 & 31\,400 & 3.9 & 0.63 & high intensity state, no spectral change\\
  \hline
\end{tabular}
\end{table}

In the following days, three more flares (flares~2 to~4, see
Fig.~\ref{fig:lc}), as listed in Table~\ref{tab:flares}, were
observed. All three flares were shorter and less intense than
flare~1 on a science window averaged basis, but still reached SCW
averaged intensities close to 1\,Crab.  Unfortunately, \integral was
in engineering mode during flare~3 and the standard OSA pipeline
rejects these data;  therefore, only count rates could be obtained
and no further analysis of this flare is possible.

On 2003 December 7 (JD\,245981.10), another intense flare was observed
(designated Flare~5, see Fig.~\ref{fig:lc}). Unlike flare~1, during
which the brightness of the source increased rapidly, it took
$\sim$8\,h for flare~5 to reach its SCW averaged maximum 20--40\,keV
flux of $\sim$1.2\,Crab. The decay lasted $\sim$5\,h until \vela
reached its pre-flare count rate of $\sim$35\,\cps (200\,mCrab in
20--40\,keV). Although quite bright, flare~5 is therefore
significantly less intense than giant flare~1, and also far longer,
i.e. it is a high intensity state. In the following, we designate
flares that have a duration of many hours and exhibit a rather slow
increase, i.e. $T_\text{rise}/T_\text{total} \gtrsim 0.5$, as high
intensity states, as opposed to flares which exhibit a rapid increase,
such as  flare~1. With $1.22\times10^{41}$\,ergs the fluence of
flare~5, however, was very similar to the energy release of flare~1.

\begin{figure}
\includegraphics[width=\columnwidth]{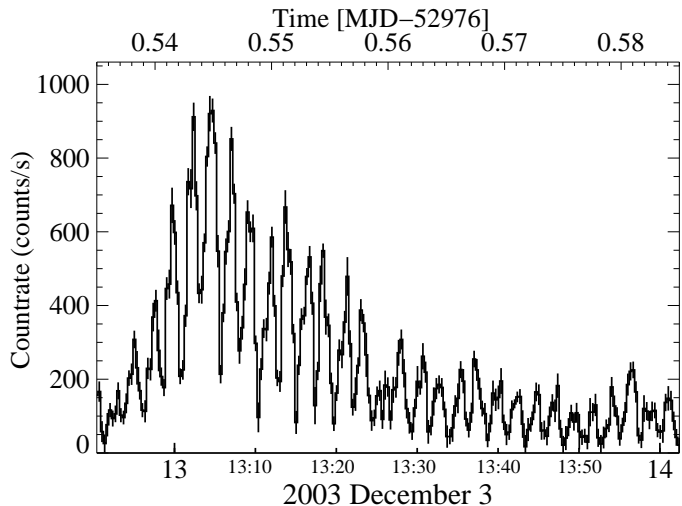}
\caption{ISGRI 20--40\,keV light curve of the short flare~3 on 2003
  December 3 with a time resolution of 20\,s.  The count rate
  increases from $<$40\,\cps at the onset of the flare to 929\,\cps
  during the peak (thus exceeding even flare~1) in less than 800\,s
  only to decay within 1200\,s to a count rate of $<$60\,\cps again.
  Since the flare is extremely short, its peak intensity is not
  evident in a light curve using SCW long bins as in
  Fig.~\ref{fig:lc}.}
\label{flare2}
\end{figure}

A more detailed analysis using a light curve with a 20\,s time
resolution showed that the source pulsated all the time, including
during flares, and exhibited the well known strong pulse-to-pulse
variations (Fig.~\ref{fig:flare_lc}). In the maximum, a peak
20--40\,keV count rate of \ca920\,\cps in one 20\,s time bin
(5.2\,Crab) was observed for flare~1 and \ca930\,\cps (5.3\,Crab) for
flare~3. Flare~3 on December 3 was therefore also a giant flare (see
Fig.~\ref{flare2}). However, flare~3 was significantly shorter: the
entire flare lasted less than 2000\,s, but it was as bright as flare~1
(see Table~\ref{tab:flares}).

\begin{table}
  \caption{Observed off states during our \vela observation. The
    duration is the time between the onset of the off state (i.e.
    sudden intensity decrease) and end of the off state (sudden
    intensity increase; see Fig.~\ref{fig:offstate}). Off state~5
    is special as the source stays at a very low luminosity level after the
    end of the off state; in this case, the duration equals the time when no
    pulsations are detected (cf.\ Sect.~\ref{disc:offstates}). 
}
\label{tab:offstates}
\begin{tabular}{crrr}
\hline
\hline
Off state & \multicolumn{1}{c}{start} & \multicolumn{1}{c}{stop} & \multicolumn{1}{c}{duration} \\
          & \multicolumn{1}{c}{[MJD]} & \multicolumn{1}{c}{[MJD]} &
          \multicolumn{1}{c}{[s]} \\
\hline
1 & 52981.095 & 52981.105 & 860 \\
2 & 52981.275 & 52981.291 & 1380 \\
3 & 52981.348 & 52981.365 & 1470 \\
4 & 52981.373 & 52981.379 & 520 \\
5 & 52981.422 & 52981.445 & 1980 \\
\hline
\end{tabular}
\end{table}

\begin{figure}
\includegraphics[width=\columnwidth]{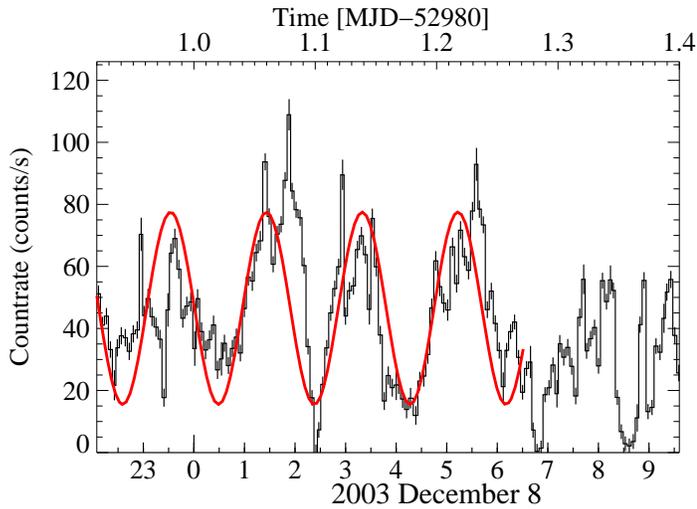}
\caption{A closeup of a part of the pulse averaged light curve of
  \vela (i.e. a light curve with a time resolution of 283\,s). A
  quasi-periodic modulation with a period of \ca6820\,s is evident in
  this part of the light curve, as indicated by the overplotted sine
  wave.  Note that during the trough between 2\,h and 3\,h, and
  especially following the quasi-periodic modulation, the count rate
  decreased several times to zero for a short time, i.e. the source
  became completely undetectable by ISGRI (see
  Table~\ref{tab:offstates} and Sect.~\ref{sect:lc} for
  discussion).}
\label{fig:qpo}
\end{figure}

Extending the analysis to the non-flaring parts of the light curve, we
detected a quasi-periodic oscillation (QPO), similar to other
accreting X-ray pulsars \citep[e.g. in 4U\,0115+63 or
V\,0332+53;][]{heindl99b,mowlavi05a,takeshima94a}.  Detecting a QPO,
especially if it is transient, can be difficult.  Using dynamic power
spectra (PSDs), we found a short-lived QPO with a period in the range
of a few 1000\,s, but no evidence for the presence of any
quasi-periodic behavior below 140\,s (half the pulse period,
Sect.~\ref{sect:period}). The short-lived QPO with a period of
\ca6820\,sec appears to be regular and inconsistent with  pure
stochastic behavior (see Fig.~\ref{fig:qpo}).  Subsequent period
searches on the corresponding data subset clearly detect the period.
We note that the quasi-periodic modulation shown in Fig.~\ref{fig:qpo}
is far stronger and inconsistent with the NOMEX effect\footnote{For
  an explanation of the NOMEX support structure and its implications,
  see e.g. \citet{lubinski04a} and \citet{reglero01a}.}, which can
cause intensity variations from SCW to SCW, but not within a given
SCW.

In this case, the amplitude of the oscillations was large, i.e. the
mean count rate was \ca45\,cps with an amplitude of 25\,cps.
During the bottom part of the oscillation, however, the
flux sometimes decreased to approximately zero counts for a short time (see
Fig.~\ref{fig:qpo}), i.e. the source turns off (see the following
discussion).

Although these modulations are evident (see Fig.~\ref{fig:qpo}),
the statistics for the entire observation are weak since this
event was short lived. Furthermore, it is well known that pure
red noise  can also produce quasi-periodic flux variations \citep[for a
discussion using \object{Mrk\,766} as an example, see][]{benlloch01a},
therefore these modulations must be treated with care.

\begin{figure}
\centerline{\includegraphics[width=\columnwidth]{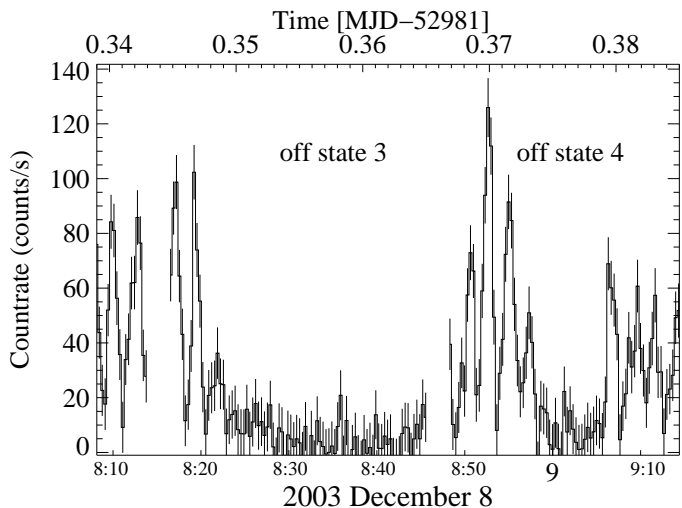}}
\caption{20--40\,keV light curve with 20\,s resolution of off states~3
  and~4 of \vela, which are most remarkable as the source suddenly
  becomes undetectable by ISGRI and then turns on again two times
  within one hour. During the off state no pulsations are discernible,
  while outside the off state the double pulse-profile of \vela is
  visible. }
\label{fig:offstate}
\end{figure}

During and after the part of the light curve in which the
quasi-periodic oscillation was present, we observed several off
states, during which no significant residual flux was detectable by
ISGRI and no modulation due to the otherwise omnipresent pulsations
was present (Fig.~\ref{fig:offstate}).  The onset of these off states
occured without any identifiable transition phase.  The source simply
switches off, or more precisely, the luminosity of the source drops
below the detection limit of ISGRI. The same statement is also true
for the end of the off states, where \vela switches on again and
immediately resumes its normal (pre off state) intensity level and
behavior.  We identified five such off states during the entire
observation, which all occured in the 12\,h from MJD\,52981.0 and
MJD\,52981.5 (Table~\ref{tab:offstates}).  Figure~\ref{fig:offstate}
shows off states~3 and~4, which are separated by only 600\,s (about
two and a half pulse periods). In between the two off states, \vela
pulsated normally and at a normal intensity level. Since the two off
states occured in rapid succession, they could also be considered as
one longer off state interrupted by 2 normal pulses.

Off state~5 differed from the other four off states: the onset of off
state~5 is also sudden, but after the pulsations are observed
again, the flux level remains low for several thousand seconds.  Off
states~1 to~4 are shown in Fig.~\ref{fig:qpo}, where the flux declines
to almost zero for a short time (note, however, that the light curve
in Fig.~\ref{fig:qpo} has a different binning).

\subsection{Pulse Period}
\label{sect:period}
Due to strong pulse-to-pulse variations, the determination of the
period of pulsars with long pulse periods such as \vela is
rather difficult. We therefore used a two step approach. First,
using epoch folding \citep{leahy83b}, we derived an approximate period of
$283.5\pm0.1$\,s for the \vela observation. We then used this 
period as a starting point for a pulse-profile template-fitting
approach. In this approach we derive sets of pulse-profiles from the
beginning to the end of the observation using sufficient data for
the long-term pulse-profile to emerge. Since the long-term pulse-profile
is known to be extremely stable \citep[see
e.g.][]{staubert80a,raubenheimer90a,kreykenbohm02b}, it was then
possible to compare these pulse-profiles and determine the shift in
seconds between any two profiles. Using these shifts, the starting
period, and the number of elapsed pulses between the two profiles,
we performed a linear fit to obtain a refined period and possibly a
$\dot P$ with the refined period until no shift was detectable between
the first profile at the start of the observation and the last profile
at the end of the observation \citep[for a detailed
explanation of this method, see][]{fritz06a}. With this technique, we were
able to derive a refined averaged pulse period of
$283.5320\pm0.0002$\,s. Our analysis, however, showed, that the pulse
period exhibited fluctuations on timescales shorter than one orbital
period with an amplitude of up to 0.003\,s (for a detailed discussion
of the evolution of the pulse period, see Staubert et al., in
preparation).  No pulse period ``glitches'' were observed, which is
consistent with the short duration of the flares, during which no
significant angular momentum transfer onto the neutron star is
expected.

\begin{figure}
\includegraphics[width=\columnwidth]{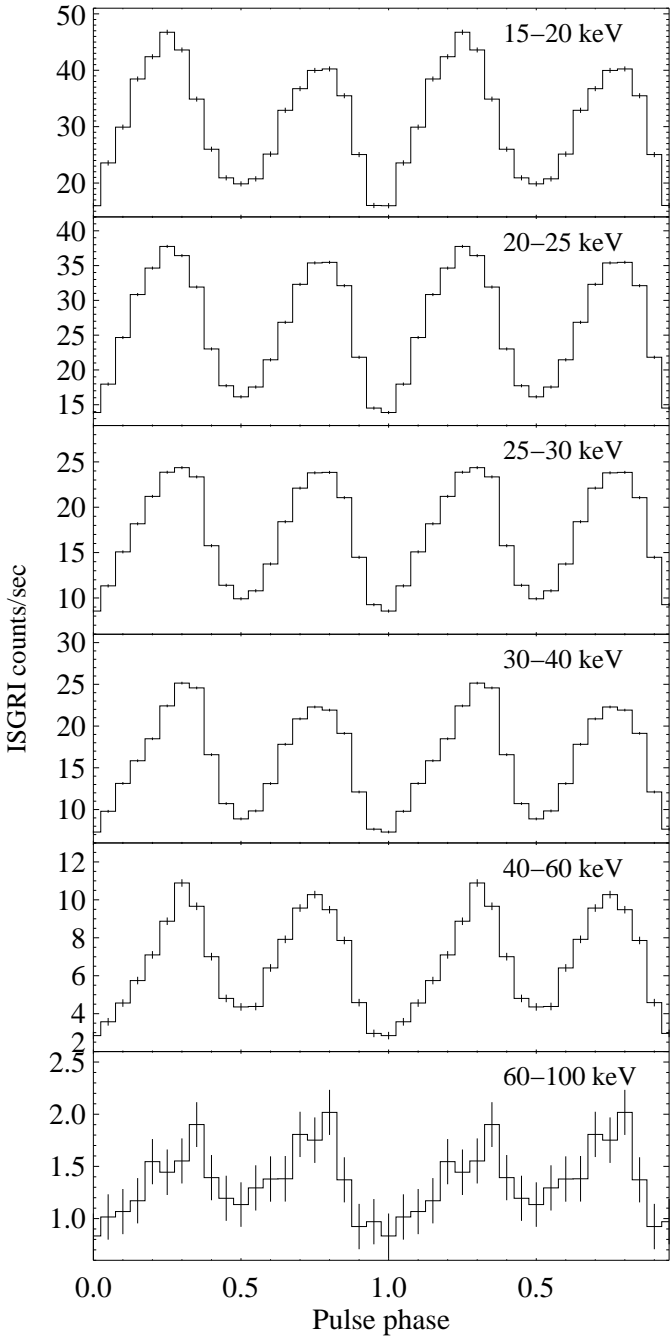}
\caption{Energy resolved pulse-profiles of \vela using all data
  outside eclipse including giant flare~1.  The folding pulse period
  is 283.5320\,s (see Sect.~\ref{sect:period}). The profile has been
  adjusted such that the pulse minimum is at phase 0. The pulses are
  clearly detected also in the highest energy band up to 100\,keV.}
\label{fig:pulseprofiles}
\end{figure}

The pulse-profile of \vela is known for its remarkable energy
dependence \citep[see for example][and references
therein]{labarbera03a,kreykenbohm02b,kreykenbohm99a,raubenheimer90a}
and, despite its complicated structure, also for its long-term
stability \citep{raubenheimer90a,staubert80a}. While the complicated
profile for energies below 10\,keV consists of up to 5 peaks, the
profile at energies $>$20\,keV consists of two simple energy
independent peaks.

The pulse-profiles shown by \citet{labarbera03a,kretschmar97c}, and
\citet{staubert80a}, however, showed that the high-energy pulse
profile of \vela is also quite complex: around 30\,keV, the main peak
shows a triangular shape, i.e. a linear rising flank followed by a
sudden and sharp fall while the secondary peak is sinusoidal
\citep[Fig.~3 of][]{kretschmar97c}.  Pulse-profiles were obtained for
all data including the flares, but without the eclipses (see
Fig.~\ref{fig:pulseprofiles}). The \integral data confirm these
previous results and show that the secondary peak remains sinusoidal
from 15\,keV up to 100\,keV, while the main peak evolves from a 
simple almost sinusoidal shape around 20\,keV into a ``triangular''
shape with a smooth linearly rising flank followed by a steeply
falling edge at 40\,keV and above. Apart from confirming earlier
results, our \integral data therefore indicate that the high-energy pulse-profile is
stable on timescales of decades and over significant luminosity ranges.

We also derived pulse-profiles for flare~1 only. Despite the dramatic
intensity increase, however, the number of pulses added was rather
small. The signal-to-noise ratio diminished, and significant
systematic artefacts became evident in the folded light curve due to
the low number of pulses. In any case, no significant change in the
pulse-profiles was evident within the uncertainties. We therefore do
not discuss these profiles further here.

\subsection{The eclipse}\label{sect:eclipse}

\begin{figure}
\centerline{\includegraphics[width=\columnwidth]{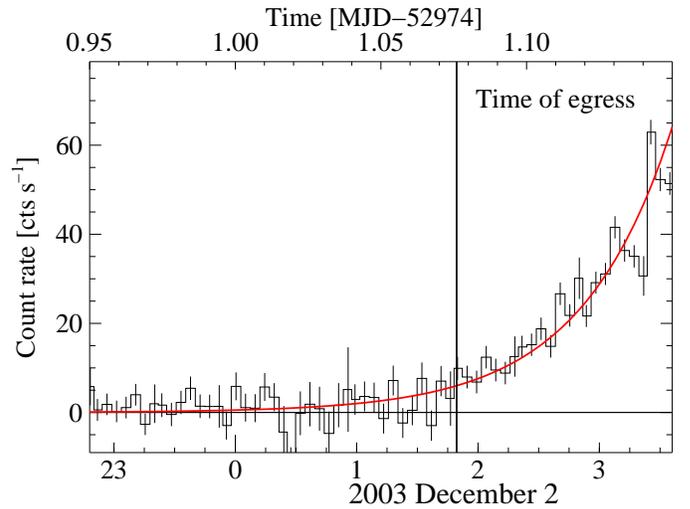}}
\caption{Close-up of the egress of \vela from the first eclipse using a
  pulse averaged light curve of \vela (time resolution of 283.5\,s). The
  flux increases exponentially with time until a normal flux level is
  reached as indicated by the solid red line. The egress time is
  determined as the time when the count rate was no longer consistent
  with zero and is marked by a thick vertical line.}
\label{fig:egress}
\end{figure}

During this long observation of \vela, two eclipses were observed.
When examining Fig.~\ref{fig:lc}, the eclipses appear to
be far longer than the 1.69$\pm$0.06\,d reported by
\citet{watson77a}. A detailed analysis of the ``technical''
circumstances, i.e. the \integral orbits and the perigee
passages of the satellite, revealed that the perigee passages of
\integral occur directly before both observed eclipses of \vela (see
Fig.~\ref{fig:lc}). Fortunately, there are a few science windows
between the end of the perigee passage (i.e. the start of a new
revolution) and the ingress into eclipse, such that it is possible to
determine the start and end of the eclipse accurately.
Since the optical companion of \vela is a massive B0.5Ib super giant
with an extended variable atmosphere \citep{sato86a}, it is difficult
to determine the ingress and egress times of \vela in soft X-ray
observations to high precision due to strong photoelectric
absorption \citep[see e.g.][and references
therein]{stuhlinger07a}. At higher energies, such as in the
20--40\,keV \integral data, photoelectric absorption poses less of a
problem, although the
determination of the egress or ingress time is only possible with an
uncertainty of several 100\,s (see Fig.~\ref{fig:egress}), due to the
changing extension of the atmosphere of the companion star and the
general variability of the source.

Using a barycentered pulse averaged light curve, we determined the
ingress and egress times of both eclipses with an accuracy of about
one pulse period (see Fig.~\ref{fig:egress}). Using these times, we
derived the duration of the eclipses in the 20--40\,keV energy band to
be $1.70\pm0.01$\,d, consistent with the 1.69\,d reported by
\citet{watson77a}.  Then we derived the center times of the two
eclipses to MJD\,52974.223 and MJD\,52983.195, respectively.  Using
these two times, we derived a new mid-eclipse reference time of
MJD\,52974.227$\pm0.007$. Note that the uncertainty in our measurement
of the mid-eclipse reference time is far smaller than for earlier
measurements \citep{vanderklis84a,watson77a}, since the time
resolution is significantly higher.  To compare this mid-eclipse time
with the $T_{90}$ of the ephemeris given by \citet{bildsten97a}, we
need to convert the mid-eclipse time $T_\text{ecl}$ to the time of
mean longitude $T_{90}$. Since the eccentricity $e$ of \vela is
non-zero, this conversion is not straightforward. The $T_{90}$
is slightly offset from the center of the eclipse by $\Delta T =
T_\text{ecl} - T_{90}$ (for a purely circular orbit, $T_{90}$ and
$T_\text{ecl}$ are identical, and see also Fig.~\ref{fig:lc}). This
offset is given by \citep{deeter87a}:
\begin{equation}
\Delta T = -\frac{P_\text{orb}}{\pi} e \cos \omega 
\left( 1 + \frac{1}{2 \tan^2 i} + \frac{(\sin i - \beta) (1 - \beta
    \sin i)}{2\beta\sin^2 i} \right)
\end{equation}
where $\beta$ depends on the semi-major axis and the stellar
radius:
\begin{equation}
\beta = \sqrt{1-\frac{(R/a)^2}{1-e^2}} .
\end{equation}
For the case of \vela, $\beta = 0.8$ \citep{deeter87a}.  Applying the
corresponding values from Table~\ref{tab:ephemeris}, we obtain a
significant offset of 0.226$\pm$0.005\,d (the uncertainty is due to
the uncertainty in the inclination $i$). We therefore calculate
MJD\,52974.001$\pm$0.012 to be the new measurement of $T_{90}$ (in
good agreement with the $T_{90}$ = 52974.008$\pm$0.008 obtained from
pulse-timing analysis by Staubert et al., in prep.).  Comparing this
$T_{90}$ with the ephemeris of \citet{bildsten97a}, we find that the
time of mean longitude $T_{90}$ differs by about 0.005\,d. This shift
is unsurprising given that the uncertainty in the orbital
period is of the order of 0.00004\,d and the uncertainty in
$T_{90}$ of \citet{bildsten97a} was 0.0012\,d.  Since then, \vela has
orbited its companion 455 times such that the accumulated uncertainty
amounts to a maximum shift of about 0.02\,d, far more than the
observed shift of 0.005\,d.  We therefore provide an updated $T_{90}$
and orbital period $P_\text{orb}$ for the ephemeris in
Table~\ref{tab:ephemeris}, but leave the other parameters of the
ephemeris unchanged.

\begin{table}
  \caption{Ephemeris of \citet{kerkwijk95a} and
    \citet{bildsten97a}, which have been used
    for obtaining the pulse period and the improved mid eclipse time
    $T_\text{ecl}$, the corresponding $T_{90}$, and the improved
      orbital period $P_\text{orb}$. See text for discussion of the
    new $T_{90}$.
} 
\footnotetext{XXX}
\label{tab:ephemeris}
\begin{tabular}{lr@{\,}ll}
  \hline
  \hline
  $T_{90}$ & MJD\,44278.5466&$\pm0.0037$ & (1) \\
  $T_{90}$ & MJD\,48895.2186&$\pm0.0012$ & (2) \\
  $T_\text{ecl}$ & MJD\,52974.227&$\pm0.007$ & (3)\\
  $T_{90}$ & MJD\,52974.001 &$\pm0.012$ & (3) \\
  $P_\text{orb}$ & 8.964416 &$\pm0.000049$\,d & (1) \\
  $P_\text{orb}$ & 8.964368 & $\pm0.000040$\,d & (2) \\
  $P_\text{orb}$ & 8.964357 & $\pm0.000029$\,d & (3) \\
  $a \sin i$    & 113.89 & lt-sec & (2) \\
  $i$ & $>73$& $^\circ$ & (1) \\
  Ecc. $e$ & 0.0898 & $\pm0.0012$ & (2) \\
  $\omega$ & 152.59 & $\pm0.92$ & (3) \\
\hline
\end{tabular}

References. (1)~\cite{kerkwijk95a}; (2) \cite{bildsten97a}; (3) this work
\end{table}

\subsection{Spectrum}
We combined all ISGRI data outside the eclipse to create a spectrum of
high signal-to-noise ratio. Since a fully working physical model of
the X-ray production mechanism in accreting X-ray pulsars does not
exist due to the complexity of the problem \citep[significant progress
has, however, been made; see e.g.][]{becker07a}, an empirical model
has to be used, which consists of a power law modified
by a high-energy cutoff at high energies. However,
special care has to be taken not to introduce any artificial features
by a ``break'' or a sudden onset of the cutoff as in the case of the
``High-energy cutoff'' in \textsl{XSPEC} \citep[see for example ][and
references therein]{kreykenbohm99a}. This is true particularly if the
source exhibits cyclotron resonant scattering features (CRSFs).  We
therefore used a power law modified by the Fermi-Dirac cutoff
\citep{tanaka86a} to model the spectrum. This model 
describes well the continuum of \vela in the ISGRI range, i.e. above
20\,keV, and was in fact used successfully in the past to fit the spectrum
of \vela \citep{kreykenbohm99a} and other sources. To model the well
known CRSF in the spectrum of \vela at 50\,keV \citep{kendziorra92a},
we used a Gaussian optical depth profile
\citep[GABS,][]{coburn02a}. The full model for the ISGRI spectrum is
then given by
\begin{equation}
I_\text{cont}(E) \propto E^{-\Gamma} \times \frac{1}{
\exp\left(\frac{E-E_\text{cut}}{E_\text{fold}}\right)+1} \times \exp\left(-\tau_{\text{GABS}}(E)\right)
\end{equation}
A systematic error of 1\% was applied to account for the uncertainties
in the response matrix of ISGRI.

\begin{figure}
\includegraphics[width=\columnwidth]{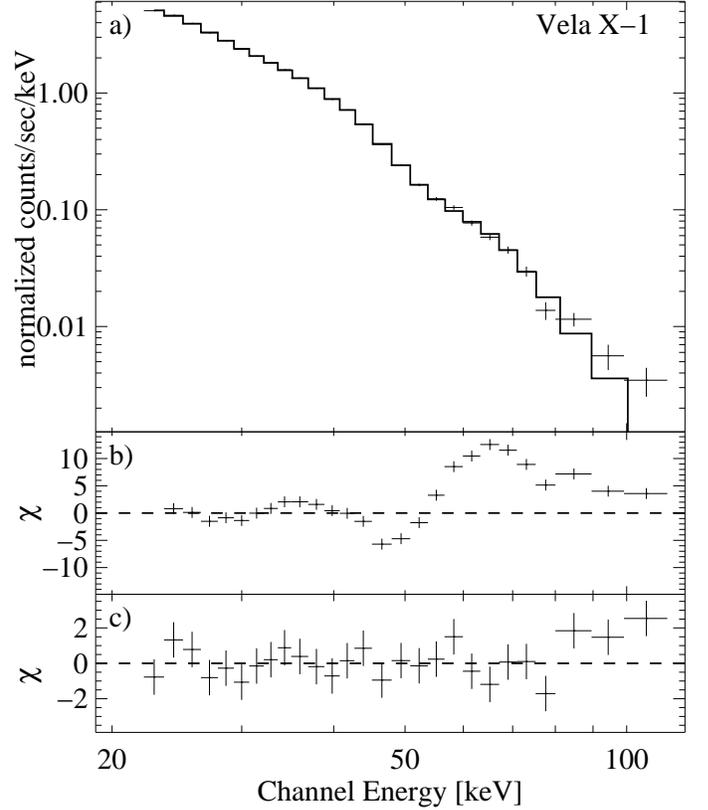}
\caption{Spectrum of \vela as obtained by \integral-ISGRI \textbf{a}
  data and folded model. \textbf{b} residuals when using only a power
  law modified by the Fermi-Dirac cutoff. \textbf{c} residuals for the
  best-fit model with a cyclotron absorption line at
  53.4\err{0.7}{0.6}\,keV. See Table~\ref{tab:spectrum} for all
  spectral parameters.}
\label{fig:spectrum}
\end{figure}

\begin{table}
  \caption{Fit results for the overall spectrum excluding the
    eclipse. When fixing $\Gamma$ to one of the values found in the literature
    \citep[fixed~1 and fixed~2;][]{kreykenbohm99a,labarbera03a}, \ecut and \efold are better constrained,
    but the resulting \redchi-values are higher.
    All uncertainties here 
    and elsewhere in the paper are on a 90\% confidence level. The DOF
    are the degrees of freedom.}
\label{tab:spectrum}
\begin{tabular}{lrr@{}cr@{}lr@{}l}
\hline
\hline
\multicolumn{2}{c}{Parameter} & \multicolumn{2}{c}{free} &
\multicolumn{2}{c}{fixed 1} &\multicolumn{2}{c}{fixed 2} \\
Exposure & [ksec] & \multicolumn{2}{c}{508} & \multicolumn{2}{c}{508} & \multicolumn{2}{c}{508}\\
\hline
$\Gamma$ && 1.6 & \err{0.3}{0.6} & 1.8 &\ fix & 2.0 &\ fix\\
E$_\text{cut}$ & [keV] & 35.6 & \err{7.5}{11.5} & 41.3 & \err{1.8}{1.5}& 47.3 & \err{3.5}{0.6}\\
E$_\text{fold}$ & [keV] & 11.2 & \err{0.5}{0.3} & 10.9 &
\err{0.4}{0.3} & 10.0 & \err{0.6}{0.6}\\
E$_\text{C} $ & [keV]& 53.4 & \err{0.7}{0.6} & 53.6 & \err{0.6}{0.6}
& 53.5 & \err{0.9}{0.3}\\
$\sigma_\text{C}$ & [keV] & 7.6 & \err{0.7}{0.7} & 8.0 & \err{0.6}{0.5}
& 8.2 & \err{0.9}{0.2}\\
$\tau_\text{C} $ & &  1.0 & \err{0.1}{0.1} & 1.0 & \err{0.1}{0.1}
& 1.1 & \err{0.2}{0.1}\\
\hline
\redchi (DOF) & & 1.3 &\ (20) & 1.4 &\ (21) & 1.6 &\ (21) \\


\end{tabular}
\end{table}

The broad band continuum is described accurately by a power law with
Fermi Dirac cutoff, but the spectral parameters were not well
constrained (see Table~\ref{tab:spectrum}). This was expected, since
no data below 20\,keV were available to enable us to determine the
spectral slope, the cutoff, and folding energy simultaneously,
especially since the cutoff energy is expected to be around
\ca20\,keV, close to the lower end of the energy range. Fixing
$\Gamma$ to some (arbitrary) value from the literature allows to
$E_\text{cut}$ and $E_\text{fold}$ to be constrained more accurately,
however, depending on the choice of $\Gamma$ also results in a rather
bad fit (see Table~\ref{tab:spectrum}). In any case, the parameters of
the CRSF do not depend strongly on the choice of $\Gamma$.

After fitting the broadband continuum, highly significant features
remain, which are due to the cyclotron line at \ca50\,keV (see
Fig.~\ref{fig:spectrum}b).  After the inclusion of a Gaussian
absorption line at 53.4\err{0.7}{0.6}\,keV, the resulting fit is
acceptable ($\redchi = 1.3$ with 20 degrees of freedom) and no
significant features remain (see Fig.~\ref{fig:spectrum} and
Table~\ref{tab:spectrum}).  Note that the shallow line like residuals
below \ca30\,keV in Fig.~\ref{fig:spectrum}b are \emph{not} due to the
disputed CRSF at 25\,keV but merely a consequence of  the
50\,keV line.  The question of whether the 25\,keV line exists or not
cannot be answered by this observation because no data below 20\,keV are
available, which would be crucial to determine the continuum and
to detect a CRSF at 25\,keV\footnote{Using \integral data from a 2\,Msec long observation in 
  \integral AO3 including JEM-X data, \citep{schanne07a},
  however, confirmed the existence of the
  25\,keV CRSF.}.  The resulting best fit is shown in
Fig.~\ref{fig:spectrum}a and the residuals in
Fig~\ref{fig:spectrum}c. Note that after the inclusion of the Gaussian
absorption line at 53.4\,keV, some residuals remain (see
Fig.~\ref{fig:spectrum}c) at 80\,keV. In this energy range, some strong
background lines are present in the ISGRI background \citep[tungsten
and lead,][]{terrier03a} that might be responsible for these
residuals. However, it is well known that CRSFs have a  complex
line shape \citep{schoenherr07a,araya00a,araya99a} and are therefore
not well modeled by simple Gaussian or Lorentzian functions \citep[see for
example][]{kreykenbohm05a,heindl99b}. The residuals present in
Fig.~\ref{fig:spectrum}c could therefore also be due to an insufficient
description of the CRSF by a Gaussian absorption line and therefore
incorrect modeling of the underlying continuum.  To improve this
unsatisfactory situation and to derive real physical parameters from
the line shapes, efforts are made to create line shapes using Monte
Carlo simulations \citep{schoenherr07a}.

A more detailed analysis of the spectrum using various spectral
models, studying the evolution of the spectral parameters with time,
and, in particular, using phase resolved spectroscopy, and a comparison
with previous work is beyond the scope of this paper and will be
discussed in a forthcoming publication. 

\subsection{Spectral evolution during the flares}
\label{Sect:hardness}

After studying the averaged spectrum, we now consider the spectral
evolution of the source during the flares, especially giant
flare~1. Figure~\ref{fig:lc}b shows the hardness ratio over the flare,
defined to be
\begin{equation}\label{eq:hardness}
\text{HR} = \frac{H-S}{H+S}
\end{equation}
where $H$ is the count rate in the hard band (40--60\,keV) and $S$ the
count rate in the soft band (20--30\,keV). The hardness ratio remained
constant throughout most of the observation, i.e. no correlation with
orbital phase was evident. During the flares, the behavior of the
hardness ratio, however, was remarkable. Shortly
before the onset of giant flare~1 as well as during the flare, a clear
deviation in the hardness ratio from its overall average value of
$-0.74$ was evident; the hardness ratio declined suddenly to $-0.82$
and later during the flare even to $-0.85$ (see Fig.~\ref{fig:lc}b). A
similar behavior was observed for flare~4: with the sudden onset of
the flare, the hardness ratio dropped from a pre-flare value of
approximately $-0.72$ to $-0.84$, the same level as in giant flare~1,
although flare~4 was far shorter and reached only a third of the peak
flux of flare~1.  This result means that the source became
significantly softer during these flares. Strikingly, however, this
softening did \emph{not} always occur: during flares~2 and~5, no
significant change in the hardness ratio was apparent. In fact, the
hardness ratio during these flares remained constant at the average
value.  This result is remarkable, since with a duration of \ca9\,h,
flare~5 lasted far longer than the other flares and its peak flux
reached $>$50\% of the peak flux of giant flare~1; it was therefore
far longer and brighter than flare~4, but it is associated with no
spectral softening.

Although the source was bright during flares~1 and~5, it was not
possible to derive meaningful spectral fits for spectra from both
flares since the net exposure time was still short. We refrain from
discussing the spectra of the flares in more detail since the
spectral parameters could not be reliably constrained.

\begin{figure}
\includegraphics[width=1.0\columnwidth]{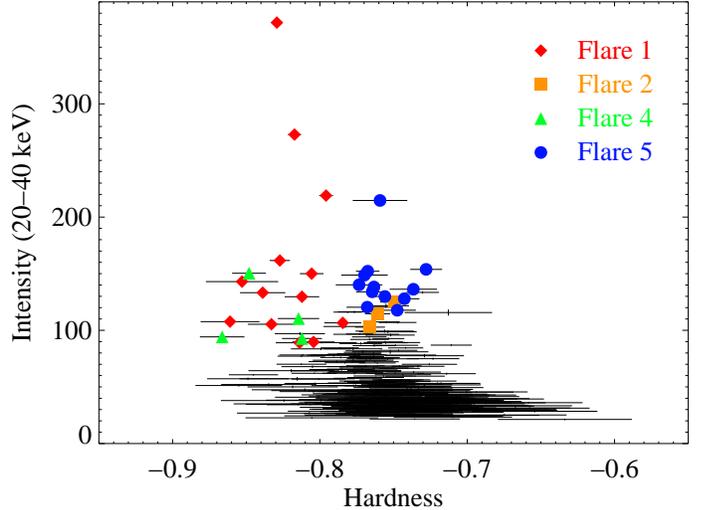}
\caption{Hardness Intensity Diagram of \vela; data from the eclipses
  have been excluded. The time resolution is one science window
  (typically 1800\,s). The hardness ratio is defined as given in
  Eq.~\ref{eq:hardness}.  The data points from the flares (see
  Fig.~\ref{fig:lc} and Table~\ref{tab:flares}) are indicated by
  individual symbols. Note that flares~1 and~4 both show significant
  softening opposite to flares~2 and~5, which show no softening (see
  Sect.~\ref{Sect:hardness}). No data are available for flare~3 (see
  Sect.~\ref{sect:lc}). }
\label{fig:hid}
\end{figure}

``Hardness intensity diagrams'' (HIDs) are a  useful
  way to study the spectral evolution of a source. A HID
for the full observation of \vela, excluding the eclipses, is shown in
Fig.~\ref{fig:hid}. As expected, the HID of \vela does not show the
\textsf{q} shape observed for black holes \citep{fender04a}. In \vela,
most of the data points are clustered around the average values of
intensity and hardness ratio. The only exception are the data
points from giant flare~1 and the other flares, which are above the
general cluster of data points due to their -- by definition -- higher
intensity. The data points of flare~1 and flare~4 are 
shifted, due to the spectral softening during the flare, as is already
evident from the hardness ratio (see Fig.~\ref{fig:lc}b). Note that
the softening did not evolve during the flares, but the flares are
softer than the average spectrum from the very onset.  On the other
hand, flares~2 and~5, although of comparable intensity, did \emph{not}
show any softening. Instead, the data points of the flares form two
distributions (flares~1 and~4 versus flares~2 and~5) that are
disjunct (see Fig.~\ref{fig:hid}). On the whole, the HID shows a
slight hardening towards lower count rates.

Another method of searching for spectral evolution are color-color
diagrams (CCDs); e.g. V\,0332+53 shows an interesting evolution
\citep{reig06a}.  We derived a CCD for \vela but no interesting
behavior was found.

\section{Discussion}
\label{sect:discussion}

\vela has been known for a long time to have a highly time-variable
light curve and to show intensity variations of up to a multiple or a
fraction of the original intensity on time scales of days, hours, or
even seconds.

\subsection{The flares}

Although \vela has exhibited extensive flaring activity in the past
\citep[see][among
others]{kreykenbohm99a,haberl94a,lapshov92a,haberl90a,nagase83a},
however, giant flares (as flares~1 and~3) had not been seen
before. The first question is therefore whether these flares are a
rare phenomenon or just could not be detected.  The problem is that
all these flares are rather short -- even giant flare~1 and long
flare~5 have a duration of a few hours.  Unless the source is
constantly observed as in our observation, typical sky monitors such
as the All Sky Monitor (\asm) onboard \xte fail to detect these
flares. The \xte-\asm monitors the entire observable sky regularly,
typically observing \vela up to 10 times a day, but often only two or
three 90\,s long dwells per day are available\footnote{See the
  \xte-\asm project web page at \url{http://xte.mit.edu/}.}. This
monitoring (which is furthermore irregular) does not allow us to
detect such short-lived events as these giant flares.  As shown by
\citet{staubert04a}, the \xte-\asm failed to detect the giant flares~1
and~3 when only taking 90\,s dwells into account. During flares~2
and~5, it showed only minor increases, but given the typical
uncertainty in the \asm data and the overall scatter, no unusual
behavior of \vela could be detected. We note that flare~1 was present
when relaxing the assumption of using only 90\,s ASM dwells to dwells
with $>$80\,s exposure; however, these dwells often have larger
$\chi^2_{\rm red}$-values than recommended. Furthermore, detecting
short-term variability with the \asm is problematic, since the \asm
sometimes also erroneously detects significant flux (even at the level
of the flares) while \vela is in eclipse. We therefore conclude that
monitoring flares on a single dwell level with the \asm is unreliable
and uninterrupted observations are required. This finding also readily
explains why these giant flares have to date been unnoticed, since
long uninterrupted observations spanning at least one full binary
orbit of \vela are only rarely performed.

The analysis of the hardness ratio of the flares in
Sect.~\ref{Sect:hardness} shows that there seem to be at least two
different types of flares (see also Fig.~\ref{fig:hid}): the first
type (flares~1 and ~4) showed dramatic increases in the count rate and
the onset of the flare was very sudden.  These flares showed significant
spectral softening during the flare. They could appear at any time,
i.e. were unrelated to the previous evolution in a similar way to
flare~1, which is superimposed on a downward trend. Although flare~4
was short, the change in the spectral hardness ratio during the flare
was significant (see Fig.~\ref{fig:lc}b). The second type was more similar to
a high intensity state than an actual flare. Flares of this type are
longer than the first type. The hardness ratio does not change during
these flares, indicating that the spectrum does not change, and the
source simply becomes brighter. For the short flare~3, no hardness ratio
could be obtained since \integral was in engineering mode during that
time; but given that it is short and features a very dramatic rise, it
is likely that it belongs to the first type.

Although the hardness ratio in Fig.~\ref{fig:lc}b is already 
conclusive, a hardness ratio comparing the high-energy spectrum (above
20\,keV) with the low-energy spectrum (below 10\,keV) would be even more
interesting since changes due to photoelectric absorption are only
visible at low-energies. However, no JEM-X data are available.  It is
therefore impossible to determine how bright the flares were at lower
energies with our current data set. Given the flux level of
more than 5 Crab in the peak of flare 1 and the significant softening
observed (see Sect.~\ref{Sect:hardness}), we can assume that the
source was also extremely bright at lower energies.  However, high
photoelectric absorption could have dampened the brightness again in the
classical X-ray band.  The evolution of \nh during a
flare could possibly help to differentiate between flare types and their underlying
mechanisms, which is impossible with our current data set.

The mechanism behind these different types of flares, however, is not
understood.  Simulating asymmetric adiabatic accretion flows onto a
neutron star in the wind of OB stars, \citet{taam89a} demonstrated
that a temporary accretion disk may form in systems such as \vela. The
formation of this temporary accretion disk is the result of an
interaction between the incident flow and shocks in the wake region
and is a general property of a binary system consisting of a neutron
star and an OB companion. Another consequence is a reversal of the
accretion flow. Associated with the flow reversal is the destruction
of the accretion disk, resulting in a significantly increased
accretion rate. During this short phase, the material stored in the
temporary accretion disk is accreted onto the neutron star.
\citet{taam89a} showed that these flow reversals occur on timescales
of several hours, and predicted flares that would last from 15 to
60\,minutes, in agreement with the observation of flares~3 and~4 
and also with the 1\,h long flare
observed by \citet{kreykenbohm99a}. Furthermore, the overall flaring
recurrence timescale during our observation agrees with that given by
\citet{taam89a}.

Later hydrodynamical studies found that wind-accretion onto the
neutron star is already a highly instable process in itself. The
accretion wake following the neutron star contains dense filaments of
compressed gas with density variations of a factor of 100 compared
with the undisturbed wind. When accreted, these density fluctuations
produce abrupt changes in the X-ray luminosity \citep[see
e.g.][]{blondin90a}. Even more important for wind-accretion, however,
is the velocity structure of the wind, since
\begin{equation}
  L_X \propto \frac{\rho}{v^3}
\end{equation}
where $\rho$ is the density and $v$ the velocity of the wind
\citep{bondi44a}. Hydrodynamical simulations have shown that not only
the density but also the velocity of the stellar wind changes
dramatically with time including sharp drops and spikes \citep[see
e.g.][]{runacres02a,runacres05a}.  Furthermore, the shock trailing
the neutron star oscillates with brief periods of disk formation,
forcing the accretion flow to change its pattern, generating the
so-called ``flip-flop instability''
\citep{matsuda91a,matsuda87a}. This instability then produces
disk-like rotational inflows that change their direction repeatedly
\citep[see][and references therein]{benensohn97a}, and is an intrinsic
phenomenon that occurs whenever a gas stream flows past a neutron star
or black hole, no density or velocity gradient being necessary
\citep{matsuda91a}. The timescale for this flip-flop behavior is
calculated to be of the order of 45\,min \citep{benensohn97a}, which
would agree with the flare observed by \citet{kreykenbohm99a} and
flares~3 and~4 in Fig.~\ref{fig:lc}. On the other hand, the flip-flop
behavior does not explain flare~1, which is superimposed on a general
downward trend, because it is far longer (several hours), nor flare~5
which is about 12\,h long, since the calculated accretion rates vary
on an even shorter timescale of \ca100\,s or less \citep[see e.g.
Fig.~3 of][]{benensohn97a}.

While the above scenario explains well the short flares seen during
our observation (flares~2, 3, and~4) and the overall rapid variability
of \vela, the flip-flop instability fails to explain the intense long
flares presented in this paper (flares~1 and~5) that last several
hours; another mechanism must therefore be at work that can alter
$\dot M$ by a factor of up to 10, not only for \ca100\,s, but for many
hours. \citet{kaper93a} demonstrated that, due to the clumpiness of
the shocked wind, the local density varies by a factor of 100, which
can explain the flaring X-ray luminosity \citep[see
also][]{oskinova08a}.  \citet{leyder07a} argued that dense clumps
trapped in an otherwise thin and more homogeneous wind might be
responsible for long flares, when the clumps are being accreted. Since
giant flare~1 lasts several hours, however, it is unlikely that a
single blob could feed the accretion for such a long time, given
that the 6\,h and 12\,h durations of giant flare~1 and of flare~5
correspond to a considerable fraction of the neutron star
orbit. Instead, the clumpy OB star wind \citep{kaper93a,lucy80a} is
probably viscously smeared out in a (small) accretion disk of the
system. This filled accretion disk can then feed the neutron star with
a significantly higher $\dot M$ than usual over several hours. A
change of $\dot P$, however, cannot be observed, since the transferred
angular momentum during such a flaring episode is far too small (see
also Sect.~\ref{sect:period}). When \vela is less active, the OB star
wind is probably less structured. In summary, we conclude that the
observed long flares are due to a strongly structured OB star wind.

Since it is difficult to detect or monitor the flares unless
during a pointed observation, it is not possible to determine how
frequent such events are.  Apart from the flares in our observation,
reports of flares -- although far less intense -- are common
in the literature \citep[][among
others]{kreykenbohm99a,haberl94a,haberl90a}. In another long
(\ca2\,Msec) observation of the \vela region in 2005 November by
\integral, \vela again exhibited very intense flares
\citep{kreykenbohm06b,schanne07a}. We therefore conclude that bright
flares are quite common in \vela. However, we caution that
\vela does not always show high activity. During a long
\integral observation in Summer 2003, \vela was in a quiet phase with
only little flaring activity detected.

\subsection{Short term variability}

Although the flip-flop instability discussed in the previous section
fails to explain the large flares exhibited by \vela, it could explain
its short-term variability: \vela exhibits strong pulse-to-pulse
variations and other short-term fluctuations superimposed on the
general variability, while at the same time the long-term
pulse-profile remains constant over decades
\citep{raubenheimer90a,kreykenbohm02b}. \citet{taam91a} developed
several scenarios and the resulting behavior of the mass accretion
rate.  These authors obtain a time variation in $\dot M$ that exhibits
short flare-like events on time scales of \ca100\,s, i.e. less than
one pulse period of \vela, for an accretion rate of
$0.7\dot{M}_\text{HL}$, where $\dot{M}_\text{HL}$ is the
Hoyle-Lyttleton mass capture rate in terms of the Eddington rate
\citep{hoyle41a,hoyle39a}, and a wind velocity of about 1000\,km
$s^{-1}$; such behavior is typical for O- and B-type stars and
precisely that required to explain the observed short-term
variability.  \citet{watanabe06a} studied the stellar wind of the
\vela system and were able to explain the observed line intensities by
using a wind model developed by \citet{castor75a} with a terminal wind
velocity of 1100\,km\,s$^{-1}$, which agrees with the calculations of
\citet{taam91a}.  Since this instability and the transfer of angular
momentum appear to be intrinsic to wind-accretion \citep{matsuda91a},
the accompanying torque reversals can therefore also be expected to
exhibit randomly varying short spin-up / spin-down episodes, or a
``flip-flop'' behavior. As \citet{boynton84a} suggested, this
instability might therefore be identified with the random-walk
behavior of the pulse period, observed in many wind-accreting sources
including \vela.

\subsection{The off states}
\label{disc:offstates}

In a similar way to the flaring activity, the off states reported by
\citet{inoue84a}, \citet{kretschmar99a}, and \citet{kreykenbohm99a},
where the source was not simply weaker but below the detection limit
of the respective instruments and no pulsations were observed, are
remarkable. After the off state observed by \citet{kreykenbohm99a},
the source resumed its normal, pulsating behavior without any
transition phase \citep[see Fig.~4 of][]{kreykenbohm99a}; also no
unusual behavior of the source was observed after the end of the off
state. \citet{inoue84a} reported that these off states occur without
any prior indication. In our observation, we observed several short
off states (see Table~\ref{tab:offstates}), which also occured without
any prior indication and without a transition phase such as a slow
decay.  Furthermore, as shown in Fig.~\ref{fig:qpo}, off state~1
occurred during a phase in which the source had an average intensity
level of about 250\,mCrab. The count rate decreased dramatically to
zero for \ca850\,sec (corresponding to three pulse periods) with no
pulsations being visible. \citet{kreykenbohm99a} reported observing
\vela in an off state of 550\,sec (corresponding to 2 pulse periods)
at the start of an observation, although the total duration of this
off state is unknown, and throughout the observation, no pulsations
were detected. It is possible that the events observed by
\citet{kreykenbohm99a} and in Fig.~\ref{fig:offstate} are similar,
i.e. the observation of the former started just in the middle of a
short off state similar to that in Fig.~\ref{fig:offstate}.

The reasons for these off states and the sudden reappearance of
pulsations are not understood, but \vela appears to experience them on
a regular (but non-periodic) basis (see Fig.~\ref{fig:offstate} and
Table~\ref{tab:offstates}). Although several explanations of these
phenomena have been put forth \citep[including transiting
planets,][]{hayakawa84a}, none can explain the observed off states,
since all of these ideas (for example, variations in the mass-loss
rate of the optical companion) require a significantly longer
timescale. ``Blobs'' in the stellar wind \citep{feldmeier97a}, for
example, would not only require extremely high optical depths to
block the X-rays and gamma-rays completely, but would also need
very sharp borders to explain the observed sudden turn-off and
turn-on of the source (sometimes within a single time bin of 20\,s, as
shown in Fig.~\ref{fig:offstate}). We therefore consider it highly
unlikely that typical clumps in the stellar wind are responsible for
the off states shown in Fig.~\ref{fig:offstate} and other explanations
must be considered.

The wind of early-type super giants like \hd is known to be
inhomogeneous and clumpy \citep[see discussion above
and both][]{walter07a,blondin90a}. Typical models show that the density in
the stellar wind of super giants can vary by several orders of
magnitude \citep{runacres05a}. These density variations should not only
correspond to the presence of clumps -- i.e. regions of strongly increased density -- but
also holes -- regions of strongly reduced density. In these holes, the
density is lower than the average density of the wind by a factor of
$10^3$ \citep{runacres05a}. If the neutron star enters these holes,
$\dot M$ would then  also decrease by a factor of $\sim10^3$ and
 the X-ray luminosity would be reduced
accordingly. Furthermore, the density fluctuations predicted by these
models occur suddenly \citep[see Fig.~1 in][]{runacres05a}, as is
the onset of the off states (see Fig.~\ref{fig:lc}). On the other
hand, these models predict that the density always varies; it
could therefore be expected that the off states should be rather
common, which does not seem to be case. Therefore, some additional
mechanism must be at work that is triggered only rarely.

Another mechanism that can explain these off states is the propeller
effect \citep{illarionov75a}.  In short, the propeller effect inhibits
the accretion of material onto the compact object when the Alfv{\'e}n
radius (where the ram pressure of the infalling gas and the magnetic
pressure are equal) is larger than the co-rotation radius
\citep{pringle72a, lamb73a}. The Alfv\'en radius, however, is not
constant: it depends on the amount of infalling material, $\dot M$. If
$\dot M$ drops, the Alfv\'en radius will increase and once larger than
the co-rotation radius, accretion is no longer possible and the X-ray
source basically switches off and no pulsations are
observable. \citet{cui97b} observed this effect in \object{GX\,1$+$4}:
in very low luminosity states, no pulsations were observable while in
high luminosity states, the source was strongly pulsating
\citep[however, other explanations are also possible to explain the
absence of pulsations in GX\,1$+$4; see][and references
therein]{ferrigno07a}.  The magnetic field strength for which the
system enters the propeller regime is then given by \citep{cui97b}
\begin{equation}
B = C \times \left(\frac{P}{1\,\text{s}}\right)^{7/6}
\sqrt{\frac{F_\text{X}}{10^{-9}\,\text{erg cm}^{-2}\text{s}^{-1}}}\
\left(\frac{d}{1\,\text{kpc}}\right) \ 
\left(\frac{M}{1.4\Msun}\right)^{1/3}
\label{eq:propeller_mag}
\end{equation}
where the constant $C=4.8\ 10^{10}\,\text{G}\,$, $P$ is the spin
period of the neutron star, $F_\text{X}$ is the bolometric X-ray flux,
$d$ is the distance, and $M$ is the mass of the neutron star. Since the
strength of the magnetic field of \vela is known from the observation
of the cyclotron resonant scattering features \citep{kreykenbohm02b},
Eq.~\ref{eq:propeller_mag} can be used to obtain directly the critical
flux limit for the onset of the propeller effect:
\begin{eqnarray}
F_\text{X,Propeller} & = & 4.3 \times 10^{-7} \text{erg
  cm}^{-2}\text{s}^{-1}\nonumber \\
 & & \times \left(\frac{B}{10^{12}\,\text{G}}\right)^2
\left(\frac{P}{1\,\text{s}}\right)^{-7/3}
\left(\frac{d}{1\,\text{kpc}}\right)^{-2}
\left(\frac{M}{1.4\,\Msun}\right)^{-2/3} .
\label{eq:propeller_flux}
\end{eqnarray}
Applying the corresponding values for \vela, i.e. \hbox{$B = 2.6\cdot
10^{12}$\,G}, $P= 283.5$\,s, $d=2$\,kpc, and $M$=1.9\,\Msun (see Sect.~\ref{sect:intro}), we obtain a
critical flux of
\begin{equation*}
F_\text{X,Propeller,\vela}  \approx  1.1 \times 10^{-12}\,\text{erg cm}^{-2}\text{s}^{-1}
\end{equation*}
Compared with the typical bolometric flux of several times $10^{-9}$\,erg
cm$^{-2}$s$^{-1}$, this critical flux is lower by about three orders of
magnitude. This observed bolometric flux corresponds to an intrinsic
luminosity of $\sim6\times 10^{32}$\,erg s$^{-1}$.
We can therefore safely conclude that the propeller effect
\emph{alone} cannot force the source to switch off when \vela is in
normal accretion mode.

A possible scenario would, however, be to combine both -- individually
unsuccessful -- mechanisms: the neutron star was evidently in a region
in the stellar wind of strongly variable density, as demonstrated by the
significant overall variability of the source and the presence of many
flares. 
The models for the stellar wind predict density variations of up
to a factor of $10^{3-5}$ \citep{walter07a}.  If the neutron star
enters this region with very low density, its luminosity will drop
accordingly because the X-ray luminosity depends linearly on $\dot M$,
and produce a luminosity of the order of less than $\sim
10^{33}$\,ergs s$^{-1}$. As shown above at such low luminosity levels,
\vela enters the propeller regime. This explains why no residual
pulsations are observed during the off states.

Off state~5, which is considerably longer than the other off states
and  does not show a sudden end similar to the dip observed
by \citet{kretschmar99a}, exhibits, however, a different behavior.
Unlike the off states shown in Fig.~\ref{fig:offstate}, these dips do
\emph{not} show a very sudden onset or end; instead their onset (and
end) is rather smooth and pulsations could also be observed for some
time after the onset of the dip \citep[see Fig.~2
in][]{kretschmar99a}. These authors also observed a dramatic increase
in the photoelectric absorption (more than$10^{24}$\,cm$^{-2}$,
Compton thick). These dips therefore belong to a second class of dips
which are caused by significantly increased photoelectric absorption
\citep{charles78a} due to an optically thick cloud in the wind of the
stellar companion passing through the line-of-sight. Such clouds are
known to be quite common in super-giant systems \citep[see for
example][]{nagase86a} and also \vela \citep{watanabe06a}.  The
existence of these clouds (also referred to as blobs or clumps) in the
wind of an OB stellar companion was proposed by \citet{lucy80a},
who suggested that the winds of OB stars themselves are not
homogeneous, but break into a population of blobs that are
radiatively driven through an ambient gas. This model was later
modified to include radiatively driven shocks in the stellar
wind. Since then, models of OB star winds have usually included blobs and
shocks \citep[see e.g.][and references therein]{feldmeier97a}. During
the dip observed by \citet{kretschmar99a}, the column density, \nh, was
of the order of $10^{24}\,\text{cm}^{-2}$, similar to that measured by, for
example, \citet{leyder07a} for the density of clumps in the
wind of \object{HD 74194}/\object{IGR\,J08408$-$4503}.

We therefore conclude that off states, characterized by a 
sudden onset and end, are usually  short, and could be caused by
a sudden drop in $\dot M$ that allows \vela to enter the propeller
regime. Intensity dips, however, are significantly longer, show a
smooth transition, and spectra taken during the dip provide measurements of
photoelectric absorption of more than $10^{24}$\,cm$^{-2}$. These dips
are readily explained by a dense blob in the wind passing through the
line of sight.

\subsection{Connection with SFXTs}

The similarity between the flares and off states in \vela and the
behavior of Supergiant Fast X-ray Transients
\citep[SFXTs,][]{sguera05a} is intriguing.  SFXTs are high mass X-ray
binaries that show very brief outbursts on timescales of hours or even
only tens of minutes, and then remain undetectable for months between
outbursts \citep{negueruela08a}. The giant flares of \vela reported in
this work are similar to these outbursts, which are assumed to be due
to the accretion of a dense blob of material embedded in a thin
stellar wind \citep{walter07a}: this would then imply that
$L_\text{X}$ is a direct tracer of the density of such blobs in the
stellar wind. The same holds true for \vela, for which  giant flares such as
flare~1 are also probably due to the accretion of a dense blob in the
stellar wind, as discussed above; \citet{ferrigno08a} reached a similar
conclusion in explaining the flares observed in \object{1E\,1145.1$-$6141}.
However, we also conclude that short flares and the general
variability can be explained well by the flip-flop instability, and no
additional clouds in the stellar wind are
necessary. \citet{grebenev07a} considered, why
SFXTs are unobservable when they are not in outburst. According to
these authors, SFXTs should be bright persistent
objects since the neutron star is deeply embedded in the stellar
wind. Similar to our explanation of the off states of \vela,
\citet{grebenev07a} invoke the propeller effect to explain the absence
of detectable X-rays from the SFXTs when the sources are in
quiescence. Unlike \vela, however, SFXTs are usually in the off state,
while \vela is in a normal accretion mode. The reason why SFXTs are
switched off by default could either be a significantly thinner
stellar wind (resulting in a lower $\dot M$) or a significantly
stronger magnetic field, which would inhibit the accretion even at the
densities encountered typically in the stellar wind.  We therefore
conclude that SFXTs and \vela are very similar systems, except that
SFXTs are normally in the propeller regime, while \vela is normally in
the accreting regime. Both, however, can switch sides, i.e. SFXTs can
go into outburst, while \vela can enter the propeller regime.

\subsection{QPOs}

Several accreting X-ray pulsars show one or more quasi-periodic
oscillations (QPOs) in addition to the normal pulse period
\citep{shirakawa02a}. Accreting X-ray pulsars that exhibit long period
QPOs include \object{4U\,0115$+$63}
\citep[$P_\text{QPO}\sim500$\,s][]{heindl99b} and
\object{V\,0332$+$53}
\citep[$P_\text{QPO}\sim20$\,s][]{mowlavi05a,takeshima94a} among
others.  The reason for the existence of QPOs is not always clear,
although the 0.05\,Hz QPO in V\,0332$+$53 is thought to be due to
inhomogeneities in the accretion disk \citep{mowlavi05a}, while a
second QPO at 0.22\,Hz \citep{qu05a} is almost (but not entirely)
coincident with the pulse period of the system. It can therefore
be assumed to originate in  the X-ray production region, and 
maybe also be linked to the strong pulse-to-pulse variations in
that system.  In \object{4U\,1907$+$09}, another wind-accreting
system, a transient QPO was observed with a period of \ca18\,s
\citep{zand98a}. Due to the close similarity of this system with \vela,
we could expect to observe a QPO with a period between 10\,s and 40\,s. As
discussed above, no evidence for a QPO with a period below 140\,s was
found. However, a QPO with a long periodicity in the range of several
thousand seconds was observed instead.

The quasi-periodic behavior observed in \vela (see
Fig.~\ref{fig:qpo}) is intriguing, albeit quite short
lived.  Such a periodicity -- if real -- in the luminosity of the
source would indicate that the accretion rate onto the neutron star
also varies periodically, which implies the presence of a quasi-periodic structure
in the stellar wind of the companion star, either in density or in
velocity or both. Such periodic structures in the wind are certainly
not due to  the flip-flop instability discussed above, because the
instabilities produce a chaotic non-periodic
modulation in the accretion rate \citep[see e.g. Figs.~4 and~6
of][]{taam91a}.  These periodic structures could be driven by
instabilities in the atmospheres of very luminous stars, which are
likely to reach  distances of a few stellar radii. Various
mechanisms for the production of these instabilities are discussed in
the literature, among them radiation-pressure-driven hydrodynamical
instabilities \citep{shaviv01a,shaviv01b}, rotational instabilities
\citep{fullerton97a}, non-radial pulsations \citep{owocki02a}, and
surface magnetic fields \citep{ud-doula02a}.

One of the phenomena predicted by some models
\citep[e.g.][]{uddoula02b} are ``ray'' like structures in the stellar
wind due to the magnetic field of the OB star.  It can be imagined
that when the neutron star passes through these more or less periodic
ray structures in the wind, the varying mass accretion rate $\dot M$
could then lead to the quasi-periodic behavior shown in
Fig.~\ref{fig:qpo}. When the neutron star enters a region where the
stellar wind has a significantly lower density, the accretion rate
$\dot M$ could drop to a level at which \vela enters the propeller
regime (see Sect.~\ref{disc:offstates}), as occured in off state~1
(see Fig.~\ref{fig:qpo}).  This scenario is interesting because it
helps to understand the other off states in the following part of the
light curve (see Fig.~\ref{fig:qpo}), and to explain the density
variations discussed in Sect.~\ref{disc:offstates}.

The quasi-periodic modulations in the source intensity could
therefore easily be identified with similar structures in the wind. It
should then be possible to draw conclusions from the period and
observed intensity variations, about the applicable models and parameters
and the underlying physics. However, these models, as currently present
in the literature, are  general and applicable numbers cannot
readily be ascertained. In principle, however, as soon as these numbers can be obtained,
the systematic and long-term probing of the stellar wind structure by
the neutron star in \vela could provide physical constraints of the
various wind models.

\section{Summary}
\label{sect:summary}

We have presented the analysis of a long,
continuous \integral observation of \vela spanning about 1.7 binary
orbits. In detail, our results are:
\begin{itemize}
\item \vela was found to be in a highly active state;
\item the observation of several flares were recorded, two of them giant flares with a
  peak brightness of more than 5\,Crab;
\item a transient QPO with a period of \ca6820\,s was observed, which
  we attributed to inhomogeneities in the stellar wind;
\item several off states were identified during which the
    source becomes undetectable by \integral and no pulsations were
  visible. The onset of the off states was very sudden, in addition to their
  end;
\item the pulse period was determined to be $P=283.5320\pm0.0004$\,s with
  no evidence for a spin-up or spin-down during the entire
  observation;
\item non-sinusoidal high-energy pulse-profiles were obtained up to
  100\,keV;
\item the spectrum exhibited the well known CRSF at 53.4\,keV, but due
  to a lack of low-energy data, the elusive line at \ca25\,keV could not
  be observed;
\item \vela exhibited two types of flares: rapidly rising flares, which
  were correlated with a spectral change, and high intensity
    states, which were longer and during which the spectrum remained
  unchanged;
\item the short flares can be explained well by the flip-flop
  instability: the predicted timescales of 15 to 60\,minutes are in perfect agreement with the
  observed duration of the short flares;
\item two different types of off states exist, which are probably
  either caused by a dense blob blocking the line-of-sight, or the
  onset of the propeller effect due to a drop in $\dot M$;
\item the similarity between \vela and SFXTs is striking: giant flares
  (in SFXTs: outbursts) are probably caused by the accretion of a
  dense blob from the stellar wind, while off states (in SFXTs:
  quiescence) are likely to be caused by the onset of the propeller effect.
\end{itemize}

\acknowledgements{We acknowledge financial support from DLR grants
  50OG9601 and 50OG0501, NASA grant NNG05GK55G, and a travel grant
  from the Deutscher Akademischer Austauschdienst. IK acknowledges the
  hospitality of the Universities of Alicante, Warwick, and
  Erlangen-N\"urnberg, and the University of California at San Diego.
  We thank the members of the pulsar team supported by the
  International Space Science Institute (ISSI) in Berne, Switzerland,
  for discussions which greatly helped shape the ideas presented in
  this paper, and ISSI itself for its hospitality. JMT acknowledges
  the support of the Spanish Ministerio de Educaci\'on y Ciencia (MEC)
  under grant PR2007-0176.  This work is based on observations with
  \integral, an ESA project with instruments and science data centre
  funded by ESA member states (especially the PI countries: Denmark,
  France, Germany, Italy, Switzerland, Spain), Czech Republic and
  Poland, and with the participation of Russia and the USA. }

\bibliographystyle{aa}
\bibliography{mnemonic,aa_abbrv,ikabbrv,velax1,div_xpuls,xpuls,cyclotron,books,roentgen,satelliten,foreign,misc}

\begin{appendix}

\section{Performance of \textsl{ii\_light}}
\label{app_ii}

\begin{figure}
\centerline{\includegraphics[width=\columnwidth]{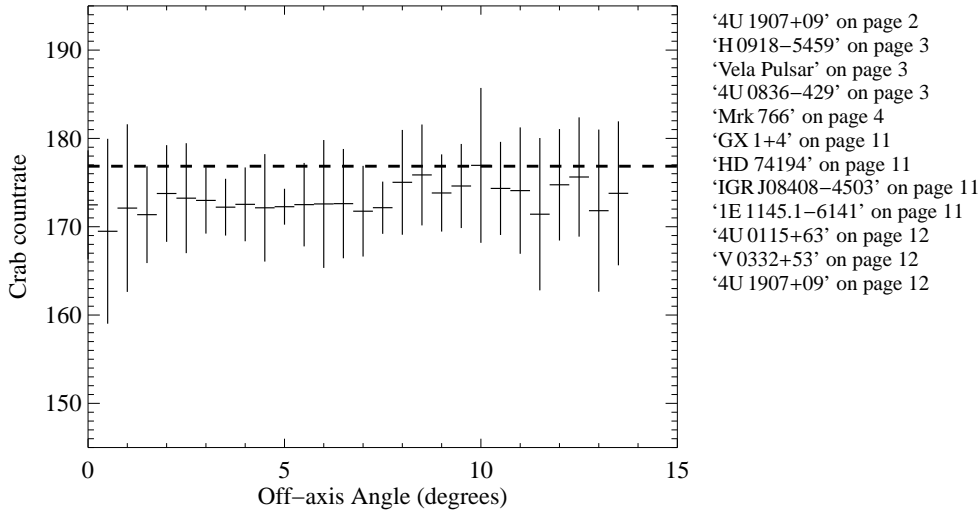}}
\caption{Dependence of the Crab count rate as measured by
  \textsl{ii\_light} on the off-axis angle. The average count rate of
  the Crab when the source is in the fully coded field-of-view
  (4\fdg{5}) as obtained by the standard IBIS pipeline is shown by the
  dashed line. Note that count rates obtained by \textsl{ii\_light}
  are within 5\% of this nominal value even up to high off-axis
  angles. \textsl{ii\_light}, however, almost always slightly underestimates
  the count rate compared to the standard pipeline. }
\label{Crab}
\end{figure}

For our analysis of the temporal properties of \vela, we require a bin
size of 20\,s. The standard IBIS analysis pipeline, however, can only
create light curves with a mi\-ni\-mum bin time of at least 60\,s and
typically a bin time of 100\,s is used. We therefore used the tool
\textsl{ii\_light} (part of OSA 7.0) to derive light curves, since it
is not only efficient and convenient to use but has the capability to
derive light curves with a time resolution as high as 0.1\,s.  Since
the IBIS cookbook cautions that \textsl{ii\_light} should only be used
to analyze the temporal behavior within a given science window, we
carefully verified the performance of \textsl{ii\_light} by using all
publicly available Crab data up to revolution 464, i.e. about 1500
science windows, corresponding to an exposure time of about
3\,Msec. First, we operated the standard IBIS pipeline up to the
imaging level to derive background and off-axis correction maps to
correct for the absorption due to the NOMEX effect \citep[for
explanation, see the IBIS analysis manual, and for the implications on
the calibration, see][]{lubinski04a}; no jumps in the count rate with
changing off-axis angle were therefore observed. We then used
\textsl{ii\_light} to obtain light curves.  Using the same time
resolution of 20\,s as for the \vela analysis, we derived an average
count rate for the Crab of 172.5\,\cps\ between 20\,keV and 40\,keV
when the source was in the fully coded field-of-view (max. off-axis
angle of 4\fdg{5}), which is very close to the 176.8\,\cps
obtained by the ibis pipeline for an entire science
window. Furthermore, we found that the deviation in the count rates
obtained with \textsl{ii\_light} is within five percent of the nominal
value up to off-axis angles of more than 14$^\circ$, which is of the
order of the expected uncertainty (see Fig.~\ref{Crab}). In any case,
\textsl{ii\_light} systematically underestimates the count rate (see
Fig.~\ref{Crab}), and the derived flux values can be considered lower
limits. In summary, we found that \textsl{ii\_light} produces usable
count rates up to an off-axis angle of $14^\circ$, we therefore
conclude that the usage of the \textsl{ii\_light} tool for our
purposes is well justified.

\end{appendix}

\end{document}